\documentstyle[12pt,aps,psfig]{revtex}
\input psfig.sty
\begin{document}
\global\topskip -.3truein
\def\la{\langle}
\def\ra{\rangle}
\def\lm{\lower 7pt\hbox { $\lim$} \atop
{\eta\rarrow 0}}
\def\tp{\tau_\phi}
\def\t0{\tau_{_0}}
\def\tpsi{t_\psi}
\def\H{Hamiltonian }
\def\o{{\rm o}}
\def\beq{\begin{equation}}
\def\eeq{\end{equation}}
\def\bA{{\bf A}}
\def\bx{{\bf x}}
\def\bq{{\bf q}}
\def\br{{\bf r}}
\def\bepsilon{{bf epsilon}}
\def\bkk{{\bf k}}
\title{Mesoscopic Physics and the Fundamentals of Quantum Mechanics}
\author{Yoseph Imry}
\address{Weizmann Institute of Science, Department of Condensed Matter
Physics,\\ IL-76100 Rehovot, Israel}
\date{\today}
\maketitle
\begin{abstract}
We start by reviewing some interesting results in mesoscopic physics
illustrating nontrivial insights on Quantum Mechanics. We then
review the general principles of dephasing (sometimes called "decoherence")
of Quantum-Mechanical interference by coupling to the environment 
degrees of freedom. A particular recent example of dephasing by
a current-carrying (nonequilibrium) system is then discussed in some detail.
This system is itself a manifestly Quantum Mechanical one and
this is another illustration of detection without the need for
"classical observers" etc.
We conclude by describing briefly a recent problem having to do with
the orbital magnetic response of conduction electrons (another
manifestly Quantum Mechanical property):
The magnetic response of a normal layer (N) coating a
superconducting cylinder (S). Some recent very intriguing experimental 
results on a giant paramagnetic component of this response are explained
using special states in the normal layer.
It is hoped that these discussions illustrate not only the vitality and interest 
of mesoscopic physics\cite{book} but also its extreme relevance to fundamental
issues in Quantum Mechanics.
\end{abstract}

\newpage
\section{Introduction}
\label{Introduction}
\par
Mesoscopic Physics \cite{book} deals with the realm which is in-between the 
microscopic (atomic and molecular) scale and the macroscopic one.  As such,
it can give us very fundamental information on the crossover between
the microscopic, purely quantum, regime and the macroscopic regime.
Macroscopic-type  elecrical measurements can be sensitive to nontrivial 
quantum phenomena in mesoscopic samples at low temperatures. 
The latter are needed to preserve the phase coherence of the electrons,
as will be discussed later.
\par
\begin{figure}
\centerline{\psfig{figure=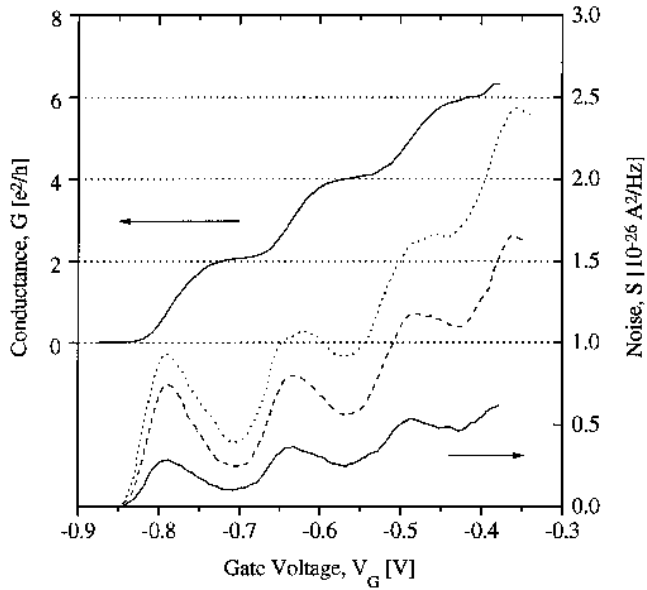,width=10cm}}
\smallskip
\caption[]{The conductance of a QPC 
in a 2D electron gas as a function of gate voltage, from ref.\cite{Rez}.
The RHS scale is for the lower curves giving the noise power for  transport 
voltages
of  1, 2 and 3 mV. } 
\label{Misha}
\end{figure}

Following the discovery of the quantum Hall effect\cite{Von} it became clear that electrical 
quantities such as conductances can display discrete "quantized" values. The basic unit being
$e^2/h$, the product of the fine-structure constant and a classical conductance related to the 
well-known impedance of free space. In fact, from the  deep and very useful Landauer\cite{Lan}
relationship between conductance and transmission, one can obtain\cite{Imr} nontrivial predictions on the 
conductance of very small orifices connecting large electron reservoirs. This conductance, 
in the above units,
is limited by the "number of quantum channels" (or transverse states below the Fermi energy).
In special cases (ballistic and tapered shapes of the orifice) the integer values can be approached.
This has been found experimentaly\cite {Qua} for carefully prepared "Quantum Point Contacts"
(QPC) in the 2D electron gas at low temeratures. 
For a 2D electron gas, the Quantum Point Contact is created by depositing, 
using lithographic techniques, two rather narrow metallic "gates' that almost touch each 
other at their tips. A negative bias depletes the electron gas beneath the gates, 
leaving a small opening,
which serves as the orifice and is controlled by the gate voltage. We show in fig. \ref{Misha} a recent 
example \cite {Rez} depicting  the conductance of such a  Quantum Point Contact as function of the gate voltage.
When the latter becomes less negative the orifice opens up and its conductance exhibits a tendency for 
plateaus at integer multiples of $e^2/h$ as shown in the figure.
\newpage
We present this particular experiment 
although flatter plateaus have been obtained before, because of a new feature: It obtained
the spectral density of the noise power due to the current through the device (as shown). 
This nonequilibrium 
"shot noise" is due to the discreteness of the electronic charge. 
It has been predicted\cite{Les}  
to peak around the transitions among the plateaus, as in fact demonstrated by the 
cited results.  We
shall later use this result in section \ref{WP}. We remark that the shot noise is an interesting 
manifestation 
of the particle nature of the electron, although it obviously appears also in situations in 
which the wave-like properties are dominant, such as tunneling through a barrier. The 
sensitivity of the shot noise to the statistics and correlations of the particles makes it
a very interesting subject for further study.

\newpage
More recently, this effect was discovered to be of relevance in various naturally-forming 
atomic scale contacts at room temperatures.

\par
The Landauer formulation enables one to consider the conductance of a mesoscopic
ring as function of an Aharonov-Bohm (AB) flux through its opening\cite{Gef}. The
conductance is found to depend periodically on the flux with a perion $\Phi_0 = hc/e$,
in agreement with general theorems ($\Phi$ is equivalent to a phase change of
$2\pi \Phi  / \Phi_0$)
and with experiment\cite{Web}. It is intersting to note that the conductance of 
macroscopically long cylinders of mesoscopic radii was predicted theoretically\cite{AAS} 
and found experimentally\cite{Shar} to have a period of $\Phi_0/2$. 
This difference of a factor of two was found to be that between a "sample specific"
and (impurity-) ensemble-averaged behavior. The latter implies averaging over many
samples all having the same macroscopic parameters but differing in the detailed
specific placement of the impurities and defects. The long cylinder consists of  many such samples 
whose resistances are added incoherently, hence the total resistance
displays the ensemle-averaged behavior
in which the fundamental $\Phi_0$ period is averaged out.
This insight opened the way to a consideration of "mesoscopic" (sample-dependent)
fluctuations in the conductance (and possibly other physical properties). We can not discuss this
here due to length limitations. The universal sizes\cite{Alt} of the above fluctuations is nevertheless
a fascinating fundamental property of these systems.
\par
When the mesoscopic ring is isolated from its leads,  one may still enquire
what happens as a function of  the AB flux through it  at equilibrium or under adiabatic
conditions.
For free electrons, it is straightforward to see that the energy levels will depend on the flux.
Therefore so will the free energy. Since the thermodynamic equilibrium current is 
the derivative of the free energy with respect to the flux, it follows that an equilibrium 
circulating current can flow in the ring and it will be periodic in the flux, with a period $\Phi_0$,
as above.  Obviously, for this to happen, the electrons must stay coherent while going
at least once around the ring.
It has been believed that scattering of the electrons by defects, imperfect
sample edges, etc. will cause this current to dissipate and therefore the equilibrium current 
should vanish in realistic systems. One of the fundamental 
quantum mechanical notions that mesoscopic physics
helped to clarify is, in fact, that elastic scattering by defects does {\em not eliminate such
coherence effects}. The impurities present just a (random) static potential. to the electron 
Well-defined coherent wavefunctions will prevail, and energy levels will be sharp and 
flux-dependent, as long as the disorder is not strong enough to cause localization.
The resulting "persistent currents" will not decay\cite{PC}.  This initially surprising fact was 
proven experimentally\cite{PCE}. The magnitude of these currents can even be {\em larger} than what
noninteracting electrons will yield. Many-body theory is necessary to explain these magnitudes
and this is still unsolved in some instances.
The fact that static impurity scattering does
not destroy phase coherence is obviously what enables the AB oscillations in the conductance
to exist. Thus {\em elastic scattering does not cause phase incoherence}. It takes {\em inelastic}
scattering to do that\footnote{As will become  clear,  here
the term "inelastic" implies  
just changing the quantum state of the environment. It is irrelevant how much
energy is transferred in this process.
This includes zero energy transfer -- flipping the environment to a degenerate state.}. 
This will be discussed in some detail in the following two sections. We remark 
that since the strength of the inelastic scattering decreases  at low temperatures,
one may reach the regime where electrons stay coherent over the whole sample. This is 
an important condition for the flux sesitivity and persistent currents to exist. For 
many experimental configurations this means that sizes in the micron range are useful
at temperatures of a few degrees K. In the nanometer size range, quantum effects should be 
observable at room temperature.
\par
Following the brief review above of some highlights from mesoscopic physics, 
we discuss in section \ref{Deph} the general principles of dephasing 
(sometimes called "decoherence")
of Quantum-Mechanical interference by coupling to the 
degrees of freedom which do not directly participate in the interference. 
Such degrees of freedom are often referred to as the "environment".
Several ways to discuss this dephasing are mentioned and their equivalence proven.
A particular recent example of dephasing by
a current-carrying (nonequilibrium) system is then discussed in some detail
in section \ref{WP} . This example also illustrates some of the principles 
discussed in section \ref{Deph}.
This system is itself a manifestly Quantum Mechanical one and
this is a good example of detection without the need for
"classical observers" etc.
We describe, in section \ref{Mota}, a recent surprising experimental result
on the orbital magnetic response of conduction electrons (another
manifestly Quantum-Mechanical property):
The magnetic response of a normal layer (N) coating a
superconducting cylinder (S) exhibits a  very intriguing  
giant paramagnetic component at very low temperatures.
This is explained in terms of an unusually large mesoscopic effect
(on scales of a fraction of a milimeter), using special states in the normal layer. 
Some concluding remarks are made in the final section.



\section{General Theory of Quantum-Mechanical Dephasing}
\label{Deph}
\par Many of the interesting effects in mesoscopic systems are due to quantum
interference.  We already saw in section \ref{Introduction} that elastic scattering by a static 
potential does not destroy quantum interference. It only modifies and 
perhaps complicates it. It takes {\em inelastic} scattering
due to the coupling of the interfering particle to its environment, to eliminate the 
interference.  The way such a coupling modifies quantum phenomena has been 
studied for a
long time, both theoretically (\cite{FV,Cal}
and experimentally. A beautiful description of the underlying physics may be found 
in the Feynman Lectures on Physics\cite{Fey}.
The effect of the coupling to the environment may be
characterized by the ``phase-breaking" time, $\tau_\phi$, which is the
characteristic time for the interfering particle to stay phase coherent, as explained
below. 
\par Stern et al.\cite{SAI} studied the way a coupling of an interfering particle
affects a two-wave interference experiment. Our discussion will be based on their
work. Two descriptions have been used of the way the interaction of a quantum system
with its environment might suppress quantum interference. The first regards the
environment as measuring the path (sometimes referred to as "Which Path" detection)
of the interfering particle. When the environment
has the information on that path, no interference is obtained. The second description
answers the question naturally raised by the first: How does the interfering particle
``know", when the interference is examined, that the environment has identified its
path? This question is answered by the observation that the interaction of a partial
wave with its environment can induce an uncertainty in this wave's phase (what counts
physically is the uncertainty of the {\it relative} phases of the paths). This may be
described as averaging out the interference pattern by turning it   into a sum 
of many patterns, shifted
relative to one another. These two descriptions were proven to be equivalent, and this
has been applied to the dephasing by electromagnetic fluctuations in metals, and by
photon modes in thermal and coherent states. Here we will review the two
descriptions, and briefly summarize the dephasing by the electron-electron
interaction in conducting matter. 
\par As a guiding example, we consider an Aharonov-Bohm
(A-B) interference experiment on a ring as described in the previous section. 
This experiment starts by a construction of two electronic wave packets, $\ell(x)$ and
$r(x)$ ($\ell,r$ stand for left, right), crossing the ring 
along its two
opposite sides. We assume that the two wave packets follow well defined classical
paths, $x_\ell(t), x_r(t)$ along the arms of the ring. The interference is examined
after each of the two wave packets had traversed half of the ring's circumference.
Therefore, the initial wavefunction of the electron (whose coordinate is $x$) and
the environment (whose wavefunction and set of coordinates are respectively denoted by
$\chi$ and $\eta$) is: 
\beq \psi(t=0)=\big[\ell(x)+r(x)\big ]\otimes\chi_{_0}(\eta)\label{3.1}\eeq
This kind of interference gives rise to h/e oscillations of the conductance.
At time $\tau_{_0}$, when
the interference is examined, the wave function is, in general,
\beq \psi(\tau_{_0})=l(x,\tau_{_0})\otimes\chi_{_\ell}(\eta,\t0)\ +\
r(x,\tau_{_0})\otimes\chi_{_r}(\eta,\t0)\label{3.2}\eeq
and the interference term is
$2\ Re\ \Big [\ell^*(x,\tau_{_0})r(x,\tau_{_0})\int\!
d\eta\chi_{_\ell}^*(\eta,\t0)\chi_{_r}(\eta,\t0)\Big ]$
Had there been no environment present in the experiment, the interference term would
have been just $2Re[\ell^*(x,\tau_{_0})r(x,\tau_{_0})]$. So, the effect of the
interaction is to multiply the interference term by
\begin{equation}
\int\! d\eta\chi_{_\ell}^*(\eta)
\chi_{_r}(\eta) 
\label{red}
\end{equation}
at $\t0$. 
This is so since the environment is not observed in the
interference experiment, its coordinate is therefore integrated upon, i.e., the
scalar product of the two environmental states at $\t0$ is taken.  The first way to
understand the dephasing is seen directly from this expression, which is the scalar
product of the two environment states, coupled to the two partial waves, at $\t0$. At
t=0 these two states are identical. During the time of the experiment, each partial
wave has its own interaction with the environment, and therefore the two states
evolving in time become different. When the two states of the environment become
orthogonal, the final state of the environment identifies the path the electron took.
Quantum interference, which is the result of an uncertainty in this path, is then
lost. Thus, the phase breaking time, $\tau_{_\phi}$, is the time in which the two
interfering partial waves shift the environment into states orthogonal to each other,
i.e., when the environment has the information on the path the electron
takes.\footnote{The question of whether somebody comes in to observe the change 
of state of the environment, does {\it not} arise.
The  discussions of the effect of that further 
observation  are  irrelevant.}  
\par The second explanation for the loss of quantum interference regards it from the
point of view of how the environment affects the partial waves, rather than how the
waves affect the environment. It is well known that when a static potential V(x) is
exerted on one of the partial waves, this wave accumulates a phase  
\beq \phi=-\int\! V(x(t))dt/\hbar\label{3.4}\eeq
and the interference term is multiplied by $e^{i\phi}$.
``A static potential" here is a potential which is a function of the particle's
coordinate and momentum only, and does not involve any other degrees of freedom. For
a given particle's path, the value of a static potential is well defined. When $V$ is
not static, but created by environment degree(s) of freedom, V becomes an operator.
Thus its value is not well defined any more. The uncertainty in this value results
from the quantum uncertainty in the state of the environment. Therefore, $\phi$ is
not definite as well. In fact, $\phi$ becomes a statistical variable, described by a
distribution function $P(\phi)$. (For the details of this description see ref\cite{SAI}). 
The effect of the environment on the interference is then to multiply the
interference term by the average value of $e^{i\phi}$, i.e.,
$\langle e^{i\phi}\rangle=\int P(\phi)e^{i\phi}d\phi$
The averaging is done on the interference ``screen". Since
$e^{i\phi}$ is periodic in $\phi$, $\la e^{i\phi}\ra$ tends to zero when $P(\phi)$ is
slowly varying over a region much larger than one period, of $2\pi$. When this
happens, one may say that the interference screen shows a superposition of many
interference patterns, mutually cancelling each other. Hence, the phase-breaking time
is also the time in which the uncertainty in the phase becomes of the order of the
interference periodicity. This is the second explanation for the loss of quantum interference.
\par The statement of equivalence between the two explanations is given by the
equation,
\beq \langle
e^{i\phi}\rangle=\int d\eta\chi_{_\ell}^*(\eta)\chi_{_r}(\eta)\label{3.6}\eeq  
When the environment measures the path taken by the particle (by $\chi_{_\ell}$
becoming orthogonal to $\chi_{_r})$, it induces a phase shift whose uncertainty is
of the order of $2\pi$. The equivalence embodied in eq.\ref{3.6} is proven as follows: 
\par We start considering dephasing of the right-hand path $\chi_{_r}$ only. The
generalization to two paths will be seen later. The \H of the environment will be
denoted by $H_{env}(\eta,p_\eta)$, while the interaction term is 
$V(\chi_{_r}(t),\eta)$ (the left partial wave does not interact with the
environment). Starting with the initial wave function (eq. \ref{3.1}) the wave function at
time  $\t0$, using  the potential in the
interaction picture $V_I(t)\equiv
e^{iH_{env}t}V(\chi_{_r}(t),\eta)e^{-iH_{env}t}$,  is
\begin{eqnarray}
\psi(\t0)\equiv \ell(\t0)\otimes
e^{-iH_{env}\t0/\hbar}\chi_{_0}(\eta)\nonumber\\ 
+r(\t0)\otimes e^{-iH_{env}\t0/\hbar}\hat T\nonumber\\
\times \exp\left[-i\int\limits^{\t0}_0\!{dt\over\hbar}
V_I(x_{_r}(t),t)\right]\chi_{_0}(\eta)\, .
\label{3.8}\end{eqnarray} where $\hat T$ is the time-ordering operator.
Hence the interference term is
multiplied by 
$\la\chi_{_0}|\hat T\exp\left[-{i\over\hbar}\int\limits^{\t0}_0\!dt
V_I(x_r(t),t)\right]|\chi_{_0}(\eta)\ra$
The interpretation of this expression in terms of a
scalar product of two environment states at time $\t0$ is obvious. The interpretation
in terms of phase uncertainty emerges from the observation that the last expression  
is the
expectation value of a unitary operator. As all unitary operators, this operator can
be expressed as the exponential of an Hermitian operator $\phi$, i.e.,
\beq \la\chi_{_0}|\hat T\exp\left[-{i\over\hbar}\int\limits^{\t0}_0 dt
V_I(x_r(t),t)\right]|\chi_{_0}\ra =\la\chi_{_0}|e^{i\phi}|\chi_{_0}\ra\,
.\label{3.10}\eeq
Hence the effect of the interaction with the environment is to
multiply the interference term by $\la e^{i\phi}\ra$, where the averaging is done
with respect to the phase probability distribution, as determined by the
environmental state
$\chi_{_0}$.
\par The concept of the phase operator $\phi$ would seem to need more clarification. 
In fact, this is only a way of describing what happens. The whole physics is contained in  
the "phasor" unitary operator $exp(i\phi)$. \footnote{$exp(i\phi)$, in distinction to the phase 
itself, has the important feature of a  built in $2\pi$ periodicity.}
The decrease of the absolute value of its average from unity determines the deterioration of the 
interference.  Nevertheless a physical interpretation of the phase operator can be obtained
in  the case where the
potentials exerted by the environment at different points along the particle's path
commute, i.e.
\beq [V_I(x_r(t),t), V_I(x_r(t'),t')]=0\label{3.11}\eeq
Then,
$\phi=-{1\over\hbar}\int\limits^{\t0}_\o\!dt V_I(x_r(t),t)$. In this case
$\dot\phi$, the rate of accumulation of the phase, is just the local potential acting
on the interfering particle, independent of earlier interactions of the particle with
the environment. One should distinguish here between two limits: for
$\la\delta\phi^2\ra\ll 1$, the environment's potential can be approximated by a 
single-particle (possibly time-dependent) potential
$V_I(x_r(t),t)\ra=\la\chi_{_0}|V_I(x_r(t),t)|\chi_{_0}\ra$,
For $\la\delta\phi^2\ra\gg 1$, on the other hand, the interference term tends to
zero. The crossover between the two regimes occurs  at
$\la\delta\phi^2\ra
\sim 1$.
\par The condition in eq. (\ref{3.11}) is typically valid when the excitation created by
the electron at one time can not be absorbed at a later time. 
This is the case for the example of an electron interacting with a free electromagnetic
field.
Also, in large many-body environments the
potential exerted by the environment on the interfering particle is usually
practically independent of the particle's history since the environment's memory time
is very short. Therefore eq. \ref{3.11} can be assumed to hold.
\par We thus see that the loss of interference due to an interaction with a dynamical
environment can be understood in the two ways discussed. The interference is
destroyed either when the state of the environment coupled to the right wave is
orthogonal to that coupled to left wave, or, alternatively, when the width of the
phase distribution function exceeds a magnitude of order unity. The interaction with
the dynamical environment turns the phase into a statistical variable, and this,
together with the fact that the phase is defined only over a range of $2\pi$,
determines the conditions for the phase to become completely uncertain. If the
potential exerted by the environment on the interfering particle at a given point
along its path is assumed to be independent of the path, the phase uncertainty is
given by,
\begin{eqnarray}\la\delta\phi^2\ra&=&\int\limits_0^{\t0}{dt\over\hbar}\int\limits_0^{\t0}
{dt'\over\hbar}\left[\la V_I(x_r(t),t) V_I(x_r(t'),t')\ra\right.\nonumber\\ 
&-&\la V_I(x_r(t),t)\ra\times \la V_I(x_r(t'),t')\ra\Big].
\label{3.18} \end{eqnarray}
This relationship will be used in the next section.
\par The exact behaviour of the interference term for $\la\delta\phi^2\ra\gg 1$, i.e.,
the value of $\langle e^{i\phi}\rangle$ for broad distribution functions, depends on
the phase distribution, $P(\phi)$. However, the description of the phase as a
statistical variable enables us, under appropriate conditions, to apply the central
limit theorem, and conclude that $P(\phi)$ is a normal distribution. The central limit
theorem is applicable, for example, when the phase is accumulated in a series of
uncorrelated events (e.g., by a series of scattering events by different,
non-interacting, scatterers), or, more generally, whenever the potential-potential
correlation function decays to zero with a characteristic decay time much shorter
than the duration of the experiment. In particular, the central limit theorem is
usually applicable for coupling to a heat-bath. For a normal distribution,
\beq \langle e^{i\phi}\rangle=
e^{i\la\phi\ra-(1/2)\la\delta\phi^2\ra}\label{3.19}\eeq This expression is exact
for the model of an environment composed of harmonic oscillators with a linear 
coupling to the
interfering waves. This model was proven in recent years to be
very useful in the investigation of the effect of the environment on quantum phenomena
(e.g. refs.\cite{FV,Cal}). 
\par It is seen from the above discussion that  
the phase uncertainty remains constant when the
interfering wave does not interact with the environment. Thus, if a trace is left by
a partial wave on its environment, this trace cannot be wiped out after the
interaction is over. Neither internal interactions of the environment, nor a
deliberate application of a classical force on it, can reduce back the phase
uncertainty after the interaction with the environment is over. This statement can be
proven also from the point of view of the change the interfering wave induces in its
environment. This proof follows simply from unitarity. The scalar product of two
states that evolve in time under the same Hamiltonian does not change in time.
Therefore, if the state of the system (electron plus environment) after the 
electron-environment interaction took place is
$|r(t)\ra\otimes |\chi_{_{env}}^{(1)}\ra+ |\ell (t)\ra\otimes
|\chi_{_{env}}^{(2)}\ra,$
then the scalar product $\la\chi_{_{env}}^{(1)}(t)|\chi_{_{env}}^{(2)}(t)\ra$ does
not change with time. The only way to change it is by another interaction of the
electron with the same environment. Such
an interaction keeps the product
$\la\chi_{_{env}}^{(1)}(t)|\chi_{_{env}}^{(2)}(t)\ra\otimes\la r(t)|l(t)\ra$ constant,
but changes $\la\chi_{_{env}}^{(1)}(t)|\chi_{_{env}}^{(2)}(t)\ra$. The interference
will  be retrieved only if the orthogonality is transferred from the environment wave
function to the electronic wave functions which are not traced on in the experiment.
The above discussion will be quite relevant for one of the aspects of the dephasing 
problem treated in the next section.
\par So far we were  concerned with the phase $\phi = \phi_r$, accumulated by
the right hand path only. The left hand path accumulates similarly a phase
$\phi_{_\ell}$ from the interaction with the environment. The interference pattern is
governed by the {\it relative} phase $\phi_r-\phi_{_\ell}$, and it is the uncertainty
in {\it that} phase which determines the loss of quantum interference. This
uncertainty is always smaller than, or equal to, the sum of uncertainties in the two
partial waves' phases. The case of noncommuting phases will not be discussed here.
\par Often the same environment interacts with the two interfering waves. 
An interesting case is when both waves  emit the same excitation of the medium.
This radiation makes each of the partial
waves' phases uncertain, but does not alter the relative phase. 
A well-known example is that of "coherent inelastic neutron scattering" in crystals
(see e.g. ref.\cite{Kit}). This process follows from the coherent addition of the
amplitudes for the processes in which the neutron exchanges {\it the same} phonon with
{\it all} scatterers in the crystal.
\par The last example demonstrates that an exchange of energy is not a sufficient
condition for dephasing. It is also not a necessary condition for dephasing. What
is important is that the two partial waves flip the environment to {\it orthogonal}
states. It does not matter in principle that these states are degenerate. Simple
examples were given by ref.\cite{SAI}. Thus, it must be emphasized that, for
example, long-wave excitations (phonons, photons) usually do not dephase the interference.
But that is {\it not} because of their low energy but rather because they do not
influence the {\it relative} phase of the paths. An equivalent way to state this
(see refs.\cite{Fey,SAI} is that, as in the Heisenberg microscope, the radiation with wavelength
$\lambda$ can not resolve the two paths if their separation is smaller than $\lambda$. 
An example where that latter effect is crucial will be mentioned below.
\par We emphasize that dephasing may occur by coupling to a discrete or a continuous
environment. In the former case the interfering particle is more likely to ``reabsorb"
the excitation and  ``reset" the phase. In the latter case, the
excitation {\it may} move away to infinity and the loss of phase can usually be regarded
as, practically speaking, irreversible. The latter case is that of an effective ``bath"
and there are no subtleties with the definition of $\phi$ since eq. \ref{3.11} may be
assumed. We point out that in special cases it is possible, even in the continuum case,
to have a finite probability to reabsorb the created excitation and thus retain
coherence. This happens, for example, in a quantum interference model due to Holstein
for the Hall effect in insulators. 
\par
Our discussion shows that the dephasing physics is fully understood with all its
subtleties and there is really no room for further semiphilosophical discussions.
The work in mesoscopic physics has provided  quantitative examinations of this.
In many typical solid-state situations one is interested in the dephasing of, 
for example, electrons, just above the Fermi energy,
performing diffusive motion due to defects, and interacting strongly via the Coulomb
interaction with all the other electrons. This electron-electron interaction provides 
in most cases the 
dominant dephasing mechanism. It is easy then to obtain the dephasing rate from 
the strength of the  inelastic scattering of the considered electron by the electron sea, i.e.
using the trace left in the environment. A straightforward semiclassical calculation
gives for a three dimensional diffusive sample\cite{SAI}:
\begin{equation}
 \tau_{ee}^{-1}=\frac{2\pi}{\hbar\Delta}\int\limits_0^{\omega_{max}}
\frac{d\omega}{2\pi}\int\frac{d^3q}{(2\pi)^3}V_{Coul}{\rm Im}
\left(\frac{1}{\epsilon(q,\omega)}\right)\langle\rho_q^2\rangle_\omega
\left[\coth{\omega\over{2T}}-\tanh{{\omega-E}\over{2T}}\right]
\label{tee}
\end{equation}
In this equation, $V_{Coul}\equiv\frac{4\pi e^2}{q^2}$ is the 
Coulomb potential, $\epsilon(q,\omega)$ is the dielectric constant, 
$\langle\rho_q^2\rangle_\omega$ is the  matrix element squared
of the density operator $\rho_q$ between two states with an energy difference 
$\omega$, averaged over disorder configurations
(it is approximately given by 
$ \left |\la m|e^{i\bq\cdot\br}|n\ra \right |^2_{av} = 
{1\over \pi\hbar N(0)}{\rm Re}\left [{1\over{i\omega+Dq^2}}\right]$).
Here N(0) is the density of states
at the Fermi energy
and  $E$ is the  energy of the electron, measured relative to
the Fermi energy.  For a system in  the linear response regime, 
the interesting case is that of $E\approx T$.  
Eq. (\ref{tee}) can be viewed as a first-order perturbation
theory contribution to the imaginary part of the electron's self
energy, where the perturbation is the complex potential ${4\pi
e^2\over q^2\epsilon(q,\omega)}$. For conductors, 
${\rm Im}\left( {1\over\epsilon(q,\omega)}\right)
\cong{\omega\over 4\pi \sigma}$. 
Eq. (\ref{tee}) was first obtained in ref\cite{AAK} by using the effect of the 
electromagnetic fluctuations due to the electron gas on the considered electron.
The equivalence of these two points of view is guaranteed by the fluctuation-dissipation 
theorem. 
\par
For low dimensions (thin films and wires) the integrations over the 
appropriate components of {\bf q} are replaced by summations and it is found that the 
remaining integrations are infrared (small q)-divergent. A careful evaluation of the phase 
{\em difference} of two paths shows that this divergence is cured by a cutoff whose 
physical meaning is exactly that low q excitation can not distinguish paths that are separated 
in space by less than $1/q$. This is in agreement with the "Heisenberg microscope"-type 
argument\cite{Fey,SAI} 
mentioned above. Once that is done, the results are in a {\em quantitative agreement} 
with experiments.
\par
Actually, the above is easily generalizable for the dephasing by the  interaction with any
system  whose response function ${\rm Im}\left( {1\over\epsilon(q,\omega)}\right)$
is known.  In that sense,  the  physics of dephasing  is understood in principle.
The case where the "environment" is {\em  not} at equilibrium
is of fundamental 
interest since the fluctuations are not given by their well-known equilibrium values 
and the fluctuation-dissipation theorem is  applicable only for the linear transport. 
An example is provided in the next section.

\section{Dephasing by a current-carrying Quantum detector}
\label{WP}
\par The quantum point contact, alluded to in section \ref{Introduction}, can be a sensitive
detector\cite{Pepper} when biased to be on the transition between two quantized 
plateaus. It is then rather sensitive to small changes in parameters, 
for example in the 
electrostatic field nearby. This sensitivity may be used to detect the presence 
of an electron in one of the arms of an interferometer, provided the two arms 
are placed asymmetrically with respect to the QPC. Following discussions by
Gurvitz\cite{Gur}, Buks et al.\cite{Buk} performed measurements confirming 
"which path" detection by the QPC. The AB oscillations  were measured 
in a ring. On one of its arms the transmission was limited by a "quantum dot"
where the electron wave would resonate for a relatively long and  
controllable (to a degree) dwell time $\tau_d$. A QPC was placed near that arm and 
the degree of dephasing due to it (determined by  $\tau_\phi / \tau_d$) 
could be inferred from the 
strength of the AB conductance oscillations. 
\begin{figure}
\centerline{\psfig{figure=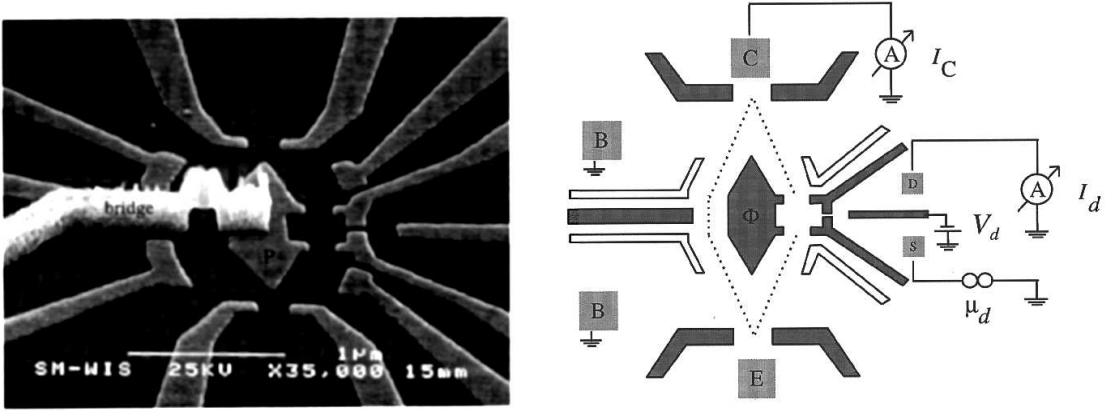,width=16cm}}
\smallskip
\caption[]{(right) A schematic view of the device used in ref.\cite{Buk}: Electrons in the 
2D gas (B) pass from the emitter (E) to the collector (C), constrained by the reflectors
(bordered empty regions). The set of gates (dark regions) deplete the electrons below them and define 
the AB ring and the quantum dot on its right arm. Near and on the right of the latter, 
the conductance ($I_D / \mu_D$) of the QPC is measured by the circuit shown.\\
(left) An SEM micrograph of the device. The gray areas are the gates 
and reflectors. An "air bridge" biases the central gate which controls the hole of the ring.}
\label{Eyalfig}
\end{figure}
Clearly, a necessary and sufficient 
condition for dephasing is that $\tau_\phi \ll \tau_d$. 
The results were in good agreement
with the theory developed by the authors\cite{Buk} and  by Levinson\cite{Lev} and,
independently, by Aleiner et al\cite{Ale}. 
We review below the main features of the theory 
for this dephasing, emphasizing the aspects that have general basic relevance.
The new interesting feature of this nonequilibrium dephasing is that a finite current is
flowing in the detector and, with increasing time,
each  electron transmitted there adds its contribution to the decrease of the overlap, eq.\ref{red}.
As discussed in the previous section, the reduction in overlap is conserved
when further thermalization of the transferred 
electrons in the downstream reservoir occurs. This is ensured by unitarity,  as long as no further
interaction with the interfering electron takes place.
The alternative picture is that   nonequilibrium (shot-noise) fluctuations
of the current in the QPC create a phase 
uncertainty for the electron in the quantum dot. While the equivalence of these two pictures is 
guaranteed by the discussion of the previous section (ref.\cite{SAI}) it is interesting and
nontrivial to see how it emerges  in detail, as we shall demostrate. This is all the more
interesting, since the former point of view by which the current fluctuations in the 
QPC cause dephasing,  seems superficially to contradict the idea behind eq. \ref{Est}. 
According to the 
latter,  dephasing appears to have to {\em  overcome} the shot-noise fluctuations, which 
therefore may be thought  to 
oppose dephasing. 
\par
In the model considered, the existence of an electron in the quantum dot
is taken to change the transmission and reflection amplitudes of the QPC
from $t$ and $r$ to $t + \Delta t$ and $r + \Delta r$. This changes the 
transmission coefficient from $T$ to $T + \Delta T$ and the conductance by
$\frac{e^2}{\pi \hbar} \Delta T$. For simplicity the QPC is taken as 
single-channel and symmetric and we consider only zero temperature
(which means practically that the temperature is much smaller than the
voltage $V$ on the QPC).
$\tau_\phi$ can be physically estimated from the condition that the 
change in the number of electrons, $\langle N\rangle = (I/e) \tau_\phi$, streaming across 
the QPC within 
$\tau_\phi$, $\langle \Delta N\rangle = \frac{e}{\pi \hbar}  V \Delta T \tau_{\phi}$
be larger than the rms fluctuations of N during the same time. For the
latter one has the quantun shot-noise result\cite{Les} according to which the mean-square fluctuation
$\langle(\Delta N)^2\rangle$ is given by $(I/e)\tau_\phi (1 - T)$ (see fig.\ref{Misha}, where the 
proportionality to $T(1 - T)$ is what generates the peaks in the noise power). 
Thus:
\begin{equation}
\frac{1}{\tau_\phi} \sim \frac{e}{\pi \hbar}\frac{(\Delta T)^2 V}{T(1-T)}
\label{Est}
\end{equation}
Below we describe three derivations, whose equivalence is
guaranteed by the discussion of section \ref{Deph}, of the precise  result
corresponding to the estimate in eq.\ref{Est}.
\par
The first two derivations consider the overlap of the 
states of the environment which are influenced by the partial waves going 
through the two arms of the ring, as in eq.\ref{red}. Ref \cite{Buk} evaluated
$\int\! d\eta\chi_{_\ell}^*(\eta)\chi_{_r}(\eta)$ where $\chi_{_\ell}$ and
$\chi_{_r}$ are the states of the QPC with the electron partial wave inside 
or outside the quantum dot (i.e. with and without the changes $\Delta t$ and
$\Delta r$ respectively). The relevant single-electron states of the QPC are the $ N(0)eV$
"scattering" states generated by the LHS reservoir. For each such outgoing state
the above overlap is easily seen to be given by $ O_1 =  r^*(r +\Delta r) + t^*( t + \Delta t)$.
For a symmetric "barrier" at the QPC, one may write the coefficients as:
\beq
r = isin\theta exp(i\phi);~~~~~~~~t = cos\theta exp(i\phi).
\label{rt}
\eeq
where $\theta$ and $\phi$ are real.
$\Delta r$ and $\Delta t$ are then parametrized in terms of $\Delta \theta$ and $\Delta \phi$ and the above overlap is given by:
\beq
O_1 = cos(\Delta \theta)  \cong 1.
\label{ov}
\eeq
In a unit time and for a length $L/2$ of the QPC arm, there are  $2v_F / L $ attempts to go 
through the QPC in each of the above states. Thus the total rate of attempts, $\nu$, 
remembering that $N(0) = \frac{L  }{2\pi \hbar v_F}$,  is given by:
\beq
\nu = \frac{eV}{\pi \hbar}. 
\label{nu}
\eeq
Thus the rate of decrease of the total overlap, i.e. the dephasing rate $1/ \tau_{\phi}$,
is $\nu$ times the single-state overlap $O_1$.  Since, from eq.\ref{rt}, 
$(\Delta T)^2 = 4(1-T)T (\Delta \theta)^2$,  one obtains:
\beq
\frac{1}{\tau_{\phi}} = \frac{eV (\Delta T)^2}{8\pi \hbar T(1-T)},
\label{Eyal}
\eeq
in agreement with the  estimate of eq.\ref{Est}.
\par
The above derivation\cite{Buk} used the overlap of the populated states in the 
QPC as modified by the interaction.
A direct way to obtain $\frac{1}{\tau_{\phi}}$ 
is to calculate the rate for real transitions induced in the QPC by the same 
interaction. This is what was esentially done in ref.\cite{Ale}. Each of the $N(0) eV$ states
populated by the higher bias (LHS) reservoir can decay into the unpopulated states
emanating from the lower bias (RHS) reservoir.  In the simplest model, the 
conductance of the single-channel QPC, of total length L, is determined  by 
a $\delta$-function potential $v \delta (x)$.  The transmission amplitude of this potential
is well-known to be $\frac{i}{i - z}$. The dimensionless parameter $z$ is defined by 
$z = \frac{v}{\hbar v_F}$.
The existence of the electron in the QD
changes $v$ to $v + \delta v$. This changes the transmission coefficient of the QPC by:
$$\delta T = -\frac{2}{\hbar v_F} T^{3/2} \sqrt{1-T}  \delta v.$$
It is easy to see that the matrix element of the perturbation $\delta v\delta (x)$
for such a transition is $$\frac{\delta v}{2L} (t + t^*) = \frac{T \delta v}{L}. $$
Evaluting this transition rate by the golden rule (but remembering that the spin in unflipped in the 
transition) and multiplying by $N(0)eV$, the number of independently decaying 
states, we arrive at the same result for 
the dephasing rate,  as in eq.\ref{Eyal}.  For the quantitative comparison,
it has to be kept in mind that the rate for real transitions is 
twice the rate of decrease of the overlap with the initial state.
\par
The  reason that the above two derivations give the same results (as was explicitly checked 
above) is that they evaluate the same overlap. One by using the states as modified by the
interaction, the other by evaluating the "real" transition rate within the unpertubed 
states. As discussed in the previous section, once the electron wave is out of the 
quantum dot and continues to diffuse along the interferometer arm with a much weaker 
interaction with the QPC, the dephasing can not be reset. This is guaranteed by unitarity,
unless further interaction of the electron wave with the detector will undo the change of
state of the latter. Before discussing below the third derivation\cite{Lev}, which uses the 
second point of view of section \ref{Deph}, we emphasize that we considered here only the 
nonequilibrium part of the dephasing (due to the finite V) . When the detector is in 
equilibrium at finite temperatures, the usual dephasing may occur. But at zero temperature 
and for an infinitesimal energy of the electron wave above the Fermi energy, no equilibrium 
dephasing takes place in usual circumstances. 
Ref\cite{Ale} pointed out that the polarization of the electron gas in 
the detector by the electron in the quantum dot causes a reduced overlap as in 
eq.\ref{red}. This is related to the 
"orthogonalization catastrophe"-type many-body effect. 
According to ref\cite{Ale} this reduction of the overlap can also exist in the present  
case at zero temperature and voltage, due
to the "Coulomb blockade" situation: The transmission through the dot is 
maximized by a gate voltage which compensates for the polarization energy of 
the electron gas.
\par
The derivation of ref\cite{Lev} utilizes the fluctuations of the phase  of the 
electron wave in the quantum dot due to the interaction with the QPC. Eq.\ref{3.18}
above expresses this phase uncertainty as an integral of the correlation function of  
this interaction.  $V_I(t)$ equals the lowest-order modulation $W(t)$ of the energy of the electron
in the quantum dot by the interaction with the QPC. Thus, the phase fluctuations increases with 
time like the integrated correlator, $K(t)$, of $W(t)$ (as in eq.\ref{3.18}). It was further 
noted in  ref.\cite{Lev}, that if $[W(t),W(t')] \not= 0$, then   $K(t)$
should be taken as the symmetrized (anticommutator) correlator. When the 
correlation time of $K(t)$, $\tau_c$,
is short (as is the case here, for further consequences, see below), one finds:
\beq
\frac{1}{\tau_{\phi}} = \frac{1}{2} \int\limits_{-\infty}^\infty K(t) dt = \pi S(0),
\label{intcor}
\eeq
where $S(\omega)$ is the power spectrum of the fluctuations of $W(t)$, i.e. the
Fourier transform of $K(t)$, $S(\omega) = 
\frac{1}{2\pi} \int\limits_{-\infty}^\infty K(t) exp(i\omega t)dt$.  
\par
A fundamentally interesting result of ref.\cite{Lev} is that $\tau_{\phi}$ also
plays the role of the decay time of the average $\hat{c}(t)$, where $\hat{c}$ is the
annihilation operator of an electron on the quantum dot. This decay was shown 
{\em not} to be related to that of the energy or the electron density of the dot.
It really reflects the decay of the coherence of the state on the dot. 
\par
Since the electron on the quantum dot couples to the QPC by changing its $r$ and $t$, it
turns out that $K(t)$ is proportional to the combination $|r^*\Delta t + t^* \Delta r|^2$ 
(which is approximately equal
to $(\Delta \theta)^2$ by virtue of eq.\ref{rt}) . A straightforward calculation 
yields for the nonequilibrium dephasing ($eV \gg k_B T$):
$$\frac{1}{\tau_{\phi}} = \frac{(\Delta \theta)^2}{4\pi\hbar} 2eV.$$
Using $(\Delta T)^2 = 4T (1 - T) (\Delta \theta)^2$, which follows from eq.\ref{rt}, this is seen to
be the same as eq.\ref{Eyal}.
\par
The correlation time of $K(t)$ is of the order of $\hbar/(eV) \ll \tau_{\phi}$. This justifies the 
assumption made in order to get eq.\ref{intcor}. The physical reason for the dephasing being 
much slower than the correlation time of the fluctuations is precisely\cite{Lev} the motional
narrowing obtained self-consistently from the same inequality.
\par
Of course, for the QPC to dephase the 
interference, the dwell time $\tau_d$ in the quantum dot has to be comparable to or larger 
than $\tau_{\phi}$. The experiments of ref.\cite{Buk} agree better than qualitatively with the above 
picture. For QPC voltages larger than thermal, the visibility of the AB interference contribution
to the conductance of the ring decreased roughly linearly in $V$ and the coefficient was in reasonable 
agreement with the above. The parameter $\Delta T$ was directly measured and the dependence on 
$T$ was qualitatively observed as well.
\section{paramagnetic orbital response of electrons in proximity to a 
superconductor}
\label{Mota}
The orbital magnetic response of conduction electrons is one of the oldest
problems in condensed-matter and statistical physics. The Bohr-van Leeuwen theorem 
ensures that this is a purely quantum phenomenon. The  persistent currents
discussed in section \ref{Introduction} and the Landau diamagnetism
are two examples of this response in normal conductors. This response is
ordinarily very small, due to the almost complete cancellation
betwen the diamagnetic and paramagnetic contributions (which must cancel exactly
in the classical limit).
In superconductors, where  the electrons condense into a single macroscopic quantum state,
the diamagnetic orbital response has the largest magnitude possible (larger by at least several 
orders of magnitude than normal persistent currents).
The intermediate situations between "normal" and "super" are therefore of interest.
\par
\begin{figure}
\centerline{\psfig{figure=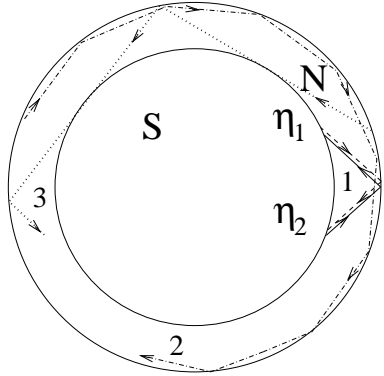,width=5cm}}
\smallskip
\caption[]{Cylindrical geometry showing electron and hole orbits.
(1) Andreev orbit. (2) ``glancing'' orbit which does not
touch the superconductor. (3) limiting case.}
\label{cylinder}
\end{figure}

The magnetic response of a normal layer (N) coating a
superconducting cylinder (S) is a good example in this connection. 
The diamagnetic response of the normal
layer (proximity effect) is related to the formation of Andreev
levels.  It turns out that usually these  levels, in which the electron
is retroreflected from the superconductor at low energies as a time-reversed hole,
do not have any paramagnetic contribution. This is the origin of the induced 
diamagnetism in the proximity layer.
At low energies, the density of such states goes linearly to
zero, which enhances the various scattering mean-free paths in the Born approximation. 
In particular, the low-energy glancing states (see fig.\ref{cylinder}), which can skip along the
outer boundary without hitting the superconductor, can have relatively large 
magnetic moments that lead
to a significant low-temperature paramagnetic correction to the
Meissner result\cite{IB}.
The order of magnitude, energy and magnetic-field
scales of this mesoscopic persistent current on an almost macroscopic scale, 
are similar to those
observed experimentally\cite{Motexp}.
\section{Concluding Remarks}
\label{conc}
\par It is hoped that the topics discussed in this paper demonstrate the value
of considering concrete examples where basic quantum issues can be addressed
by specific equations and the results eventually compared with experiment.
Mesoscopic Physics offers an ideal arena for this type of demystification. In particular,
no philsophical discussions of dephasing and "which path" detection are necessary.
Detection happens due to well-defined processes in the detector
(which is quantum mechanical by itself).
There is absolutely no need, at least in this example,  to resort to classical observers
who will look at the data and  further influence the results thereby. It is worth pointing out 
that in the solid state experiments considered here, as well as in many photon experiments,
quantities averaged over many interfering entities are considered. The probabilistic
aspects are included. What happens in an experiment with a single electron
does not have to be  considered in this type of discussion.
Even for quantities such as the shot noise (which is the fluctuations
around the average) one considers, possibly time-dependent, correlation functions, which
again are {\em averages} in the above sense.  
\par In section \ref{Mota} a new and interesting situation for orbital magnetism is reviewed.
In a system consisting of superconducting and normal components, strong correlation exist 
between electrons and holes due to Andreev reflection. These lead to the proximity-effect diamagnetism.
The same correlations decrease the low-energy density of states of these Andreev states. This,
in turn, stabilizes the special "whispering gallery" modes, which at low temperatures
give a surprisingly large paramagnetic orbital response. 
\par
The correlation (alias entanglement) due to interactions and statistics
among particles is  a ubiquitous and a very important element in condensed
matter (as well as 
in atomic and molecular) systems. Its study is very nontrivial and extremely relevant
for Mesoscopic Physics. This is one of the issues remaining in the study of the crossover 
between microscopic and macroscopic behavior.

\vspace{.3cm}
{\bf \large Acknowledgements}\\ 
The research reported here was supported by grants from the German-Israel
Foundation (GIF) and the Israel Science Foundation, Jerusalem. The author would like to thank
Y. Aharonov, C. Bruder and A. Stern for collaborations on these problems.
Ana-Celia Mota and Eyal Buks are thanked for instructive discussions on the
experiments of Refs.~\onlinecite{Motexp} and  ~\onlinecite{Buk}.  I. L. Aleiner, A. Altland, N. Argaman,
C.~W.~J. Beenakker, W. Belzig, M. Berry, E. Buks, A. Fauch\`ere, Y. Gefen,
B.~I. Halperin, D.E. Khmelnitskii, A. Krichevsky, R. Landauer, Y. Levinson, 
Y. Meir, M. Schechter, G. Sch\"on,
T.D. Schultz, A. Stern, C. Urbina, and A. Zaikin are thanked for discussions on aspects of the theory. 
E. Buks and M. Reznikov are thanked for permission to use their figures.

\end{document}